# ACOUSTIC EMISSION OF GROWING MICROCRACKS IN VIBRATION-LOADED MATERIAL

A.K. Aringazin[1, 3], N.Ya. Karasev[2], V.D. Krevchik[2, 3],
V.A. Skryabin[2], M.B. Semenov[2, 3]

[1] Eurasian National University, Astana 010008, Kazakhstan
aringazin@mail.kz

[2] Physics Department, Penza State University, Penza 440017, Russia
physics@diamond.stup.ac.ru

[3] Institute for Basic Research, P.O. Box 1577, Palm Harbor, FL 34682, USA
ibr@verizon.net

We propose mechanism describing an acoustic emission by growing microcracks in the material under external cycled load. We use the theoretical approach based on Huygens principle for elastic solid continuum with an account for dislocation creep in the zone of abrasive action. We show that the acoustic emission is anisotropic and its main direction depends not only on the effective length of microcrack but also on dynamics of dislocation structure and on diffusion processes in Kottrell zone which give rise to microcracks.

# 1. Introduction

Material failure is often accompanied by a "crackling noise". Also, acoustic emission arises prior to fracture of solid materials. In a recent review, Alava, Nukala, and Zapperi [1] emphasized that there is a considerable body of work on the relation of acoustic emission to microscopic fracture mechanisms and damage accumulation. Krylov [2] pointed out that formation of microcracks in solid materials leads to acoustic emission. Such acoustic emission is used in applications, for example, in nondestructive and express monitoring of constructions; see, e.g., monograph by Krasilnikov and Krylov [3]. Also, defects in solids are the source of acoustic emission. This enables identification of defects. Particularly, seismological methods, such as triangulation method [3], can be used here.

Since the acoustic emission reflects perhaps in the best way microfracturing dynamics, it is important to study it both in theoretical and experimental ways. Spectrum of acoustic emission can provide useful information on the dynamics of dislocations and microcracks. Also, kinetics of fracture [3] can be investigated by the use of acoustic emission data.

Krylov [2] has presented general method to analyze the acoustic emission by microcracks of arbitrary type in finite-size elastic solids. This method is based on Huygens principle. It allows one to determine both the equations of motion for microcrack edges for the material under external stress and the spectral density of acoustic emission.

However, the study of growth of microcracks made in Ref. [2] does not include the stage before microcracks appear. Mechanism which leads to the appearance of microcracks is of much importance in the case of vibration-loaded material. Such a load, as it was shown by Artemov, Krevchik, and Sumenkov [4], can stimulate the process of fretting-tiredness with a subsequent fracture of the material.

In the present paper, we propose mechanism describing the acoustic emission raised by growing microcracks in vibration-loaded materials. We refer to it as the mechanism of hidden growth of microcracks.

Physics underlying this mechanism is a dislocation creep under the action of sign-alternating stresses; see, e.g., Straw's model in the monograph by Krishtal and Mirkin [5]. As the result, a number of parallel dislocations are accumulated (clustered) close to some obstacle in the form of impurity release. In this region, a wedge-shaped cavity can therefore appear, and we consider this cavity as the source of microcracks. The acoustic emission occurs at the moment when the cavity opens. As we will show below, its spectral density essentially depends on the parameters of dislocation creep and on the loaded vibrating material.

## 2. Calculation of the effective length of wedge-shaped cavity

Let us consider the growth of "hidden" microcracks for the material under vibrating load. Oscillations in the material cause stresses with alternating sign. These stresses can both lead to motion of defects (for example, dislocations) and stimulate diffusion of impurities of doping and substitution types.

The process of vibration-stimulated diffusion for Kottrell zones in the field of dislocation deformation was considered by Artemov and Krevchik [6]. It was shown, particularly, that blurring of Kottrell zones can be accompanied by an increase of the effective length of the dislocation segment, with subsequent re-lease of dislocation out of the impurity region. Under this condition, the time interval $\tau_1$ to the dislocation release can be determined as [6]

$$\tau_1 = \tau_0 \exp(8\pi\varepsilon\varepsilon_0), \qquad (1)$$

where $\tau_0$ is the load cycle period, $\varepsilon_0 = 4r_0^2 / \ell_{c0}^2$, $r_0$ is the radius of Kottrell "cloud", $\ell_{c0}$ is the effective length of dislocation segment in the absence of vibrating load, $\varepsilon_1 = \ell^2 / \ell_d^2$, $\ell$ is the length of dislocation loop, $\ell_d$ is the diffusion length of the fixing impurity.

As shown by Kudinov [7], in the field of alternating (sign-reversing) mechanical stresses, dislocation segments with "steps" periodically pass through Kottrell zone and leave trace of vacancies after each step. Oversaturation of the vacancies in the material lead to the following increase of diffusion coefficient, $\Delta D$, of the fixing impurity in Kottrell "cloud" [7]:

$$\Delta D = D_g \frac{\eta \omega_f Q^{-1}(\sigma_0)\sigma_0}{2\pi\beta N_d E}, \qquad (2)$$

where $D_g$ is the diffusion coefficient for vacancies, $Q^{-1}(\sigma_0)$ is the internal friction coefficient, $\eta \approx 10^{-4}$, $E$ is Young modulus, $\beta$ is the nonlinearity factor for dislocation segment $(\beta \leq 1)$, $N_d$ is the density of dislocations, $\sigma_0$ is the mechanical stress amplitude, and $\omega_f$ is the vibration load frequency.

Under some critical value of stress, $\sigma_0 = \sigma_m$, for example, in the zone of abrasive action, the dislocation can go out of the impurity region with subse-quent cooperative drift in the field of alternating stresses (deformations). Under this condition, the drift speed $\vartheta_k$ is determined by the formula [4]

$$\vartheta_k = \frac{8D}{2r_0}, \qquad (3)$$

where $D = D_0 + \Delta D$, $\Delta D$ is the diffusion coefficient for fixing impurity in the absence of vibration load.

Taking into account Eq. (3) the expression for the critical velocity of creep, $\dot{\varepsilon}_k$, can be written as [4]

$$\dot{\varepsilon}_k = \frac{8D}{2r_0} b N_d, \qquad (4)$$

where $b$ is the value of Burger vector.

As it has been mentioned above, under sufficient stress value $\sigma_0$, some part of dislocations join together and form cavity near the obstacle, and the cavity is of wedge shape [5]. In the region of such clustering of $n$ dislocations, a stable cavity of the length $L = n^2 b$ can appear. Such a cavity is viewed as the reason of birth of microcrack. Supposing $n \approx \dot{\varepsilon}_k \tau_2$, where $\tau_2$ is the time to beginning of the microcrack formation, we then obtain

$$L \approx b \left[ \frac{8 b N_d}{2 r_0} (D + \Delta D) \tau_2 \right]^2, \qquad (5)$$

Here, one can use the approximate formula for $\tau_2$ obtained in Ref. [4] for some specific case,

$$\tau_2 \approx \tau_1 \exp[\alpha], \qquad (6)$$

where $\alpha = \pi\beta\tau_0/(2b\eta\omega_f C_m \Delta T \sigma_m)$, $C_m$ is the model parameter determined from experiments on the material internal friction, $\Delta T$ is the temperature difference for the material without and under vibration load.

As one can see from Eq. (5), the value of growth for the wedge-shaped cavity is determined mainly by (i) the diffusion mobility, which "decorates" the impurity dislocation, (ii) the density of dislocations, and (iii) parameters of vibration load $\sigma_0$ and $\omega_f$.

One can estimate numerically the effective length $L$ using Eq. (5). Taking the following values (see also [3]): Poisson number $\nu = 0,3$, the density of the material $\rho = 7800 \ kg/m^3$, the shear modulus $G = 8 \cdot 10^{10} \ N/m^2$, $b = 10^{-10} \ m$, $\sigma_0 = 10^5 \ N/m^2$, $2r_0 = 0,1 \ mcm$, $N_d = 10^{12} \ cm^{-2}$, $\ell = 1 \ mcm$, $D \approx 3 \cdot 10^{-14} \ cm^2/s$, $L_{co} = 0,1 \ mcm$, $\omega_f = 88 \ s^{-1}$, $T = 400 \ K$, one obtains $L \approx 0,75 \ mm$.

## 3. Calculation of acoustic emission in the regime of instant opening of the wedge-shaped cavity

Our theoretical approach to the cavity acoustics is based on Huygens principle for elastic solids [2] with an account for the dislocation creep for the material under vibrating load.

Since the cavity is viewed as an initial cut of the continuous medium under the action of alternating stresses which does not affect the effective length $L$, we can use the model of "instantly spreading" microcracks. For instant opening of the cavity with the endpoint coordinates $L/2$ and $-L/2$ under the action of external stress $\sigma(t) = \sigma_0 \theta(t)$, where $\theta(t)$ is the step function, it is reasonable approximate the function $U_z^0(x,t)$ describing dynamics of the cavity opening [2] by the following representation:

$$U_z^0(x,t) = \begin{cases} U_z(t), & |x| \leq L/2 \\ 0, & |x| > L/2 \end{cases}, \quad (7)$$

where

$$U_z(t) = \begin{cases} st, & 0 \leq t \leq \dfrac{L}{c_l} \\ \dfrac{sL}{c_l}, & t > \dfrac{L}{c_l} \end{cases},$$

$s = \sigma_0 / (\rho c_l)$, and $c_l$ is the limit velocity for the lowest symmetric Lamb mode [1].

As it has been mentioned in Ref. [2], this approximation has a quite simple physical meaning: speed of the cavity opening $s$ in this case is determined as the ratio between the external stress $\sigma_0$ (related to the abrasive action onto the surface of material) and the wave resistance of the material to the applied normal pressure $\rho c_l$.

Then, approximate expressions for spectral densities $U_r(r,\theta,\omega)$ and $U_\theta(r,\theta,\omega)$, which are valid in a long distance range, can be obtained as follows [2]:

$$U_r(r,\theta,\omega) = \frac{i\sigma_0 L^2}{4\pi^2 \rho c_l^2 \frac{\omega}{c_l} L} \sin\left(\frac{\omega}{2c_l}L\right) \frac{\sin\left(\frac{\omega}{2c_l}L\sin\theta\right)}{\frac{\omega}{2c_l}L\sin\theta} \times$$

$$\times \left(2\frac{c_t^2}{c_l^2}\sin^2\theta - 1\right)\left(-\frac{2\pi i}{\frac{\omega}{2c_l}r}\right)^{\frac{1}{2}} \exp\left[i\omega\left(\frac{r}{c_l} + \frac{L}{2c_l}\right)\right], \quad (8)$$

$$U_\theta(r,\theta,\omega) = -\frac{i\sigma_0 L^2 \exp\left[i\omega\frac{L}{2c_l}\right]}{4\pi^2 \rho c_l^2 c_t \frac{\omega}{2c_t}L} \frac{\sin\left(\frac{\omega}{2c_t}L\sin\theta\right)}{\frac{\omega}{2c_t}L\sin\theta} \times$$

$$\times \sin 2\theta \left(-\frac{2\pi i}{\frac{\omega}{c_t}r}\right)^{\frac{1}{2}} \exp\left[\frac{i\omega}{c_t}r\right], \quad (9)$$

where $r$ and $\theta$ are the polar coordinates of the observation point, $x = r\sin\theta$, $z = r\cos\theta$, and $c_t$ is the speed of the lowest SH-mode [2].

One can see from Eqs. (8) and (9) that the width of main petal in the direction diagram of acoustic emission $\Delta\theta$ both for the longitudinal $(l)$ and the transversal $(t)$ waves, $\Delta\theta_{l(t)} = 2\arcsin\left(2c_{l(t)}/(\pi\omega L)\right)$, essentially depends on the parameters of growth mechanism of wedge-shaped cavity entering Eq. (5). In Ref. [1], the effective length of microcracks was taken as an empiri-cal parameter. The present study allows one to estimate it by using Eq. (5). For example, for $L \approx 0,75\ mm$ and $\omega = 3\ MHz$, we have $\Delta\theta_l \approx 116^o$.

From Eq. (8) it follows that the radiation of longitudinal waves has maximum at $\theta = 0^o$:

$$U_r^{max}(r,\omega) = \frac{i\sigma_0 L^2 \exp\left[i\omega\frac{L}{2c_l}\right]}{4\pi^2 \rho c_l^3 c_t \frac{\omega}{2c_l} L} \sin\left(\frac{\omega L}{2c_l}\right)\sqrt{\frac{c_l}{\omega r}} \exp\left[i\left(\frac{\omega r}{c_l} + \frac{\pi}{4}\right)\right], \tag{10}$$

This equation allows one to estimate the energy density $W$ of the acoustic wave, in the approximation $\omega L/(2c_l) \square\ 1$ and $2\pi c_l/(\omega_s r) \square\ 1$. Namely,

$$W \square \frac{\sigma_0^2 L^4 \omega_s^4}{64\pi^6 \rho c_l^6}, \tag{11}$$

where $\omega_s$ is the frequency of sound wave.

Taking as an example $\sigma_0 = 10^5\ N/m^2$, $L = 0,75\ mm$, $\omega_s = 3\ MHz$, and $\rho = 7800\ kg/m^3$, we obtain the energy density $W \approx 72\ mcJ/m^3$ and the radiation intensity $I \approx 0,04\ mW/cm^2$.

## 4  Conclusions

In conclusion, we shortly formulate main results of this study. One of the possible mechanisms of the birth of microcracks is the vibration-stimulated dislocation creep, which is accompanied by the formation of the wedge-shaped cavity. Acoustic emission accompanies the moment of cavity opening. Moreover, in this case the radiation is found to be anisotropic, with the main direction being dependent on the cavity length $L$. Diffusion character of the growth dynamics of wedge-shaped cavity reveals itself by the dependence of the cavity length $L$ on the diffusion mobility of Kottrell zones.

The proposed mechanism of "hidden" growth of microcracks in vibration-loaded material is one of possible mechanisms (see, e.g., [6]), and we do not expect strict quantitative correspondence to experimental data. However, qualitative features of the acoustic emission studied in the present paper can be confronted to experiments.